\documentclass[prd, aps, superscriptaddress, preprintnumbers, twocolumn, floatfix, nofootinbib]{revtex4-1}

\usepackage{amssymb}
\usepackage{amsmath}    % need for subequations
\usepackage{graphicx}   % need for figures
\usepackage{verbatim}   % useful for program listings
\usepackage{color}      % use if color is used in text
\usepackage{subfigure}  % use for side-by-side figures
\usepackage{hyperref}   % use for hypertext links, including those to external documents and URLs
\usepackage{float}     %importanto for proper figure placement 
\usepackage{mdwlist}

\newcommand\be{\begin{equation}}
\newcommand\ba{\begin{eqnarray}}
\newcommand\ee{\end{equation}}
\newcommand\ea{\end{eqnarray}}

\begin{document}

%\title{`First Lights' from Cosmic Strings at the Cosmic Dawn}
\title{Ionization from Cosmic Strings at Cosmic Dawn}

\author{Samuel Laliberte}
\email{samuel.laliberte@mail.mcgill.ca}
\affiliation{Physics Department, McGill University, 3600 University Street, Montreal, QC, H3A 2T8, Canada}

\author{Robert Brandenberger}
\email{rhb@physics.mcgill.ca}
\affiliation{Department of Physics, McGill University, Montr\'{e}al, QC, H3A 2T8, Canada and
Institute of Theoretical Physics, ETH Z\"urich, CH-8093 Z\"urich, Switzerland}

\begin{abstract}
\noindent

Cosmic strings produce charged particles which, by emitting electromagnetic radiation, partially ionize neutral hydrogen during the dark ages.  Corrections to the ionization fraction of neutral hydrogen induced by cosmic strings could lead to new observational effects and/or new constraints on the string tension around $G\mu \sim 10^{-16} - 10^{-22}$ for values of the primordial magnetic field in the range $B_0 \sim 10^{-11}-10^{-9}$ Gauss.

\end{abstract}


\maketitle

%%%%%%%%%%%%%%%%%%
%% Introduction %%
%%%%%%%%%%%%%%%%%%

\section{Introduction}

Cosmic strings \cite{CSreviews} are linear topological defects which may have formed in the early universe.  In particle physics models which admits cosmic strings, a network of strings will inevitably form during a symmetry-breaking phase transition \cite{Kibble} and persist to late times.  The string network consists of a ``scaling system'' of ``long'' strings (strings with curvature radius larger than the Hubble horizon) and a distribution of string loops with radii smaller than the Hubble horizon. By ``scaling system'' it is meant that the statistical properties of the network are independent of time if all lengths are scaled to the Hubble horizon. In fact, the mean separation of the long strings can be shown to be comparable to the horizon. This can be seen both using analytical arguments \cite{CSreviews} and numerical simulations \cite{CSsimuls}. The scaling network of long strings is maintained by the strings intersecting each other and producing loops. The typical formation radius $R_i(t)$ at time $t$ of a string loop is again comparable to the horizon,  $R_i(t) = \alpha t$, $\alpha$ being a constant of order unity. The loops oscillate and gradually decay. For large loops gravitational wave emission \cite{GW} is the dominant decay channel, for small loops it is the process of ``cusp annihilation'' \cite{cusp} producing elementary particles which is more important.    

Cosmic strings form lines of trapped energy density. If the energy scale of the phase transition which leads to cosmic string formation is $\eta$, then the string tension (which equals the mass per unit length) is 
\be
\mu \, \simeq \, \eta^2 \, .
\ee
The dynamics of cosmic string networks can be described by the dimensionless number $G\mu$, where $G$ is Newton's gravitational constant.

The trapped energy of the strings gravitates and leads to specific signatures which can be searched for in cosmological observations (see e.g. \cite{RHBrev} for a review). The signatures of the long string network increase in strength proportional to $\mu$. Hence, searching for cosmic string signatures in the cosmological observations is a way to probe particle physics beyond the ``Standard Model'' ``from top down''. Not observing signals which strings predict can be used to rule out classes of particle physics models \cite{RHBrev2}. The current robust upper bound on the cosmic string tension is $G\mu < 1.5 \times 10^{-7}$ which comes from measurements of the angular power spectrum of the cosmic microwave background \cite{CMBconstraint}. If the distribution of string loops also achieves a scaling solution (which Nambu-Goto simulations of string evolution \cite{NGsimuls} indicate, but not all simulations based on solving the classical field theory equations \cite{Hindmarsh}), then a stronger bound of $G\mu < 10^{-10}$ can be derived from pulsar timing constraints on the amplitude of the stochastic gravitational wave background \cite{pulsar}.

According to the one-scale model of the string loop distribution, the number density of string loops per unit radius $R$ at a time $t$ after the time $t_{eq}$ of equal matter and radiation is given by \cite{CSreviews}
\be \label{loopdistrib}
n(R, t) \, \sim \, R^{-5/2} t_{eq}^{1/2} t^{-2}
\ee
for loops produced before $t_{eq}$. This distribution is valid down to a lower cutoff radius $R_c(t) \sim t$ which in the case of graviational radiation dominating the string loop decay is 
\be
R_c(t) \, = \, \gamma \beta^{-1} (G\mu) t \, ,
\ee
where $\beta$ and $\gamma$ are constants which will be introduced later.
 Hence, the loops dominate the energy density in cosmic strings, and dominate the string decay emission products.

The initial interest in the role of cosmic strings in cosmology was sparked by the idea that string loops could be the seeds for galaxies and galaxy clusters \cite{TB}. However, since a distribution of strings as the main source of cosmological perturbations does not produce acoustic oscillations in the angular power spectrum of CMB anisotropies \cite{noacoustic}, the discovery of these oscillations \cite{Boomerang} killed this idea. Nevertheless, since there is good evidence from particle physics that cosmic strings might form in the early universe, it is of great interest to look for their signatures, and in particular for effects of string loops. In fact, it was shown that string loops might yield the seeds about which high redshift supermassive black holes form \cite{BBJ}, and may play a role in globular cluster formation \cite{GC}.

An important feature of cosmic string loops is that any loop (or radius $R$) will experience a cusp at least once per oscillation time  $R$ \cite{KT}. A cusp is a region of length \cite{Olum}
\be
l_c(R) \, \sim \, R^{1/2} w^{1/2} \,
\ee
(where $w \sim \mu^{-1/2}$ is the width of the string) about a point where the two sides of the string overlap. Such a region is unstable to annihilation into a burst of particles \cite{cusp}. The primary decay products are the scalar and gauge field quanta associated with the string, but these rapidly decay into jets of long-lived elementary particles such as neutrinos, pions and electrons, and also high energy photons. The resulting flux of ultra-high energy photons and neutrinos was studied in \cite{Jane1} and \cite{Jane2}. The consequences for fast radio bursts were recently analyzed in \cite{FRB}, and the implications for the global 21cm signal were studied in \cite{Bryce}. The effects of strings typically increase in amplitude as $G\mu$ decreases, until the value of $G\mu$ for which gravitational radiation ceases to be the dominant decay channel. Thus, one can hope that searching for signals of string loops can provide constraints on lower tension strings.

Since string loops are present and undergo cusp annihilation at all times, they will contribute to early reionization of the universe. The effect of superconducting cosmic strings, which have a direct decay channel into photons, was studied in \cite{Tashiro:2012nv}. In this work, we will study the corresponding signal for ordinary (i.e. non-superconducting) strings.

After recombination, most of the hydrogen in the universe is neutral.  It is only after non-linear structure forms that hydrogen can be fully ionized again.  By analyzing the spectra of most distant quasars, one can conclude that most of the hydrogen was reionized at redshift $z \sim 6$ \cite{Pentericci:2014nia}.  However, if cosmic strings \cite{Cosmic Strings} were formed in the early universe, they could ionize the hydrogen at earlier times.
We will study how this effect depends on the string tension and on the amplitude of the primordial magnetic fields.

Throughout this paper, we use parameters for a flat $\Lambda$CDM model: The Hubble
expansion rate is taken to be $H_0=h \times 100 ~ {\rm km /s /Mpc}$ with $h=0.7$, the fractional contribution of baryons to the total energy density being $\Omega_b=0.05$ and that of matter
$\Omega_m=0.26$. The cosmological redshift $z(t)$ is related to time via 
$1+z(t) = (1 + z_{eq})\sqrt{t^{t_{eq}}/t}$ in the radiation 
dominated epoch, and $1+z(t) = (1+z_{\rm eq}) (t_{\rm eq}/t)^{2/3}$ in the 
matter dominant epoch, where $t_{eq} = (2 \sqrt{\Omega_r} H_0)^{-1}$ is the time of
equal matter and radiation and $z_{\rm eq}=\Omega_m / \Omega_r$ with 
$h^2 \Omega_r=4.18 \times 10^{-5}$. $\Omega_r$ is the current fractional contribution
of radiation to the current energy density. We also adopt natural units, 
$\hbar = c =1$.

%%%%%%%%%%%%%%%%%%%%%%%%%%%%%%%%%%%%%%%%%%%%%%
%%  Electron Spectrum of Cusp Annihilation  %%
%%%%%%%%%%%%%%%%%%%%%%%%%%%%%%%%%%%%%%%%%%%%%%

\section{Cusp Annihilation Electron Spectrum}

In this section, we compute the flux of high energy particles from cusp annihilation of non-superconducting cosmic strings. In the following section, we will then use this flux to compute the contribution of cosmic strings to ionization of the universe after the time of recombination. 

A cosmic string cusp initially decays by emitting quanta of the Higgs and gauge fields which make up the string. These particles, in turn, decay into a jet of Standard Model particles, in particular charged and neutral pions, electrons and neutrinos. The energy distribution of particles in such jets has been well studied by particle physicists, with the result that
the spectrum of charged pions produced by a cusp annihilation event is given by \cite{Jane1}
\begin{equation}
\frac{dN}{dE} \, = \,  \frac{15}{24}\frac{\mu l_c}{Q_f^2}\left(\frac{Q_f}{E}\right)^{3/2} \, ,
\end{equation}
where $Q_f$ is the energy of the primary decay quanta, which is of the order of $\eta$.
In the following we will take $Q_f \sim \eta$, leading to
\begin{equation}
\frac{dN}{dE} \, \sim \, \frac{15}{24}\mu^{1/2}R^{1/2}E^{-3/2} \ .
\end{equation}
The spectrum of electrons produced by the cusp-induced jets has the same scaling.

The energy distribution at time $t'$ (number per unit energy per unit time) of electrons emitted by cusps is found by integrating over the loop distribution at a time $t'$ from the lower cutoff radius $R_c(t')$ to the upper cutoff $R = \alpha t'$ where ($\alpha \sim 1$)
\begin{equation} 
\frac{d^2n_e(t')}{dE(t')dt'} \, = \, \int_{R_c(t')}^{\alpha t'}n(R, t')\frac{dN}{dE}\frac{1}{R}dR  \,
\label{eq:loop_dist}
\end{equation}
where $n(R, t')$ is the number density of loops of radius $R$ per unit $R$ interval at time $t'$, and the factor $1/R$ comes from the fact that there is one cusp event per oscillation time $R$.

The lower cutoff radius $R_c(t)$ is related to decay time scale
\begin{equation}
\tau \, = \, \frac{\mu L}{P} \,  = \, \frac{\mu \beta R}{P}
\end{equation}
of loops.  Here, $L$ and $R$ are, respectively,
 the string length and the loop radius and $\beta \sim 10$ is a parameter that measures how circular loops are on average (perfectly circular loops have $\beta = 2\pi$).  Usually, $P$ is the power radiated away by the dominant decay process for loops.  For example, if $G\mu > 10^{-18}$, gravitational radiation dominates during the entire time interval between recombination and the present time, so we use
\begin{equation} \label{decayGR}
P_g \, =  \, \gamma G \mu^2 ,
\end{equation}
where $\gamma \sim 100$ is dimensionless constant  \cite{GW}.  If $G\mu < 10^{-18}$, then cusp annihilation dominates as a decay mechanism in the time interval of interest, and we must use \cite{Olum}
\begin{equation} \label{decayCUSP}
P_c \, \sim \, \mu l_c/R \sim \mu^{3/4}R^{-1/2} \ .
\end{equation}
Indeed, comparing (\ref{decayGR}) and (\ref{decayCUSP}) we see that the power emission from cusp decay decreases less fast than that of gravitational radiation. Hence, for fixed time there will always be a value of $G\mu$ below which cusp decay starts to dominate.

From (\ref{eq:loop_dist}) it is clear that it is loops at the lower cutoff of the integration range which dominate the energy injected from string cusps. The reason is that the loop distribution for loops formed before equal matter and radiation scales as $R^{-5/2}$. Also, considering the $G\mu$ dependence of the electron flux from cusp annihilation, we see that the flux increases as $G\mu$ decreases as long as $R_c$ is determined by gravitational radiation. We will be interested in the values of $G\mu$ which give the largest flux. Hence, as discussed in \cite{Brandenberger:2018dfj}, both gravitational radiation and cusp annihilation are important decay mechanisms for the relevant values of $G\mu$.  Therefore, in order to obtain an accurate description of the cusp annihilation spectrum when $G\mu \sim 10^{-18}$ , we will use $P = P_g + P_c$.  This yields the following expression for the cutoff radius:
\begin{equation} \label{Req}
R_c \, = \, \left(\gamma G\mu +\mu^{-1/4}R_c^{-1/2}\right)\beta^{-1}t
\end{equation}
Analytic solutions to this expression can be found in the limits of large and small $G\mu$. In general, one should solve the equation above using numerical methods.  

In order to evaluate equation \ref{eq:loop_dist}, we use 
the loop distribution function (\ref{loopdistrib}) which holds if $R > R_c(t)$ and $t > t_{eq}$.  Since the integral in (\ref{eq:loop_dist}) is dominated by the lower cutoff, the energy distribution of electrons produced at time $t'$ can be written as
\begin{equation}
\frac{d^2n_e(t')}{dE(t')dt'} \approx \frac{5}{16} \nu \mu^{1/2}t_{eq}^{1/2}t'^{-2}R_c(t')^{-2}E^{-3/2} \ .
\label{equ:e_dist}
\end{equation}

We now wish to determine the energy distribution of electrons at some time $t$ produced by cusp decays between the time of recombination and $t$.  This is done by integrating equation (\ref{equ:e_dist}) with respect to $t'$.  The integration requires a Jacobian transformation between energies at different times to take into account the redshifting of the number density and of the energy between times $t'$ and $t$:
\begin{equation}
E^{3/2}(t')\frac{dn_e(t')}{dE(t')}  \, = \, E^{3/2}(t)\frac{dn_e(t)}{dE(t)}\left(\frac{t}{t'}\right)^{1/3}\left(\frac{t}{t'}\right)^{2} \ .
\end{equation}
Integrating with respect to $t'$, we obtain the following electron spectrum:
\begin{equation}
\frac{dn_e(t)}{dE(t)} \, \approx \, \frac{5}{16} \nu \mu^{1/2}t_{eq}^{1/2}E(t)^{-3/2}t^{-7/3}\int_{t_{rec}}^{t}dt't'^{1/3}R_c(t')^{-2} \ .
\label{eq:spec_g} 
\end{equation}
For large values of $G\mu$ when gravitational radiation dominates string loop decay, this expression scales as $(G\mu)^{-3/2}$, for small values of $G\mu$ when cusp decay is more important and when (see (\ref{Req}))
\be
R_c(t) \, \sim (G\mu)^{-1/6} \, 
\ee
the scaling of the electron energy flux is proportional to $(G\mu)^{1/6}$. Thus, the highest flux of electrons occurs in the range of $G\mu$ values when gravitational radiation of a string loop is comparable to cusp annihilation.

%%%%%%%%%%%%%%%%%%%%%%%%%%%%%%%%%%%%%%%%%%%%%%%%%%%%%%%%%
%%  Ionization Fraction of a Network of Strings Loops  %%
%%%%%%%%%%%%%%%%%%%%%%%%%%%%%%%%%%%%%%%%%%%%%%%%%%%%%%%%%

\section{Ionization Fraction of Strings}

To estimate the correction to the ionization fraction of hydrogen due to cosmic strings, will follow the approach taken in \cite{Tashiro:2012nv}.  Assuming that one photon in the frequency interval between $\omega_i$ and $\omega_f$ emitted from the string loop ionizes one hydrogen atom, we can derive the following expression for the ionization fraction due to cosmic strings: 
\begin{equation}
x(z) = \frac{1}{\alpha_rn_H(z)^2}\int_{\omega_i}^{\omega_f}d\omega \frac{d^2n_{\gamma}}{d\omega dt} \, \text{.}
\label{io_frac}
\end{equation}
where $(d^2n_{\gamma}/ (d\omega dt)$ is the number density of photons per unit frequency interval per unit time due to the string loop cusp annihilations. Here, $\alpha_r = 2.6 \times 10^{-13}$ cm$^3$/s is a recombination coefficient, $n_H(z) = 8.42 \times 10^{-6}(1+z)^3\Omega_Bh^2$ cm$^{-3}$ is the number aboundance of neutral hydrogen at redshift $z$, Photon with frequency below the Lyman-$\alpha$ frequency $\omega_i= 13.6 {\rm eV}$ do not have enough energy to ionize neutral hydrogen. Those with frequency above a cutoff frequency $\omega_f = 10^{4} {\rm eV}$  have too small a ionization cross section.

In the case of superconducting strings, there is direct photon emission from the string loops. In our case, the photons in the relevant low energy regime are mainly produced by electrons from cusp annihilation via Bremsstrahlung and Synchroton emission. There is also direct photon emission from cusps, but the corresponding spectrum is only known at high energies, i.e. for photon energies larger than the pion mass \cite{Jane1}. Hence, we focus on the more robust mechanisms of photon production in the frequency range relevant to ionization, mechanism which reply on electrons producing photons during their propagation. There are two main mechanisms - Synchrotron emission and Bremsstrahlung. Synchrotron emission depends on the strength of the primordial magnetic fields in which the string loops live, whereas Bremsstrahlung is a more general phenomenon. In the following, we will study both channels.

%%%%%%%%%%%%%%%%%%%%%%%%%%%%%%%%%
%%  Bremsstrahlung Ionization  %%
%%%%%%%%%%%%%%%%%%%%%%%%%%%%%%%%%

\subsection{Bremsstrahlung Ionization}

Charged particles emitted by a string are slowed down by the surrounding medium, which creates Bremsstrahlung radiation.  The strength of emission depends on both the flux of the charged particles produced by cusp radiation and on the density of the surrounding medium. According to \cite{Brandenberger:2018dfj,Blumenthal:1970gc}, the spectrum of photons emitted via Bremsstrahlung per unit time per unit frequency for a cusp at a given time is given by
\begin{equation}
\begin{split}
\frac{d^2n_{\gamma}(t)}{d\omega dt} \approx \left(\frac{8}{3}m_{\pi}^{-1/2}\right)\alpha_{EM}r_0^{2}K_e(t)E(t)^{-1} \left(\sum_{s}n_s(t)\tilde{\phi}_w\right) \text{.}
\end{split}
\label{p_spec1}
\end{equation}
Here, $m_{\pi}$ is the mass of the pion, $\alpha_{EM}$ is the electromagnetic fine structure constant, $r_0=m_e^{-2}$ is the Compton wavelength of the electron, $E(t)$ is the energy of the induced photons,  $n_s(t)$ is the number density of charged particles the electrons interact with and $\tilde{\phi}_{\omega} \sim 300$ is a dimensionless weak shielding coefficient.  The dependence on the flux of charged particles enters through the quantity $K_e(t)$ which is the coefficient of the power law 
\be
dn_e(t)/dE(t) \, \equiv \, K_e(t)E(t)^{-3/2} \, .
\ee
By inspection with equation \ref{eq:spec_g}, we conclude that $K_{e}(t)$ takes the value
\begin{equation}
K_e(t) \, = \, \frac{5}{16}\nu \mu^{1/2}t_{eq}^{1/2}t^{-7/3}\int_{t_{rec}}^{t}dt't'^{1/3}R_c(t')^{-2} \, \text{.}
\end{equation}
Note that we are only considering the effect of charged particles produced after the time $t_{rec}$ of recombination. Those produced before $t_{rec}$ interact with the charged plasma and rapidly lose their energy, and we will neglect their contribution. Also, the weak shielding approximation breaks down when neutral hydrogen becomes fully ionized at reionization.  Therefore, our solution only holds for times between $t_{rec}$ and the time of reionization.

\begin{figure}[h]
	\centering
	\includegraphics[width = 9.0cm]{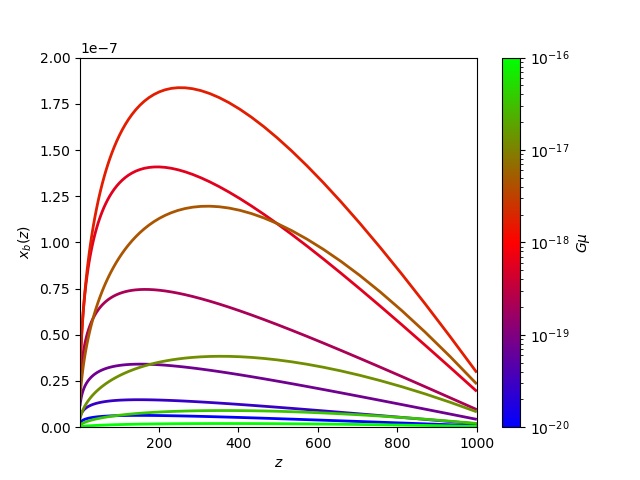}
	\caption{Ionization fraction from Bremsstrahlung at different redshifts. The values of $G\mu$ run from $10^{-19}$ to $10^{-17}$.}
	\label{fig:brem_io}
\end{figure}

To estimate the ionization fraction $x_b(z)$ for Bremsstrahlung, we  use equation (\ref{io_frac}) to obtain 
\begin{equation}
\begin{split}
x_{b}(z) \, = \, \frac{1}{\alpha_rn_H(z)^2}\left(\frac{8}{3}m_{\pi}^{-1/2}\right)\alpha_{EM}r_0^{2}K_e(t(z))\\
\times \left(\sum_{s}n_s(t(z))\tilde{\phi}_w\right)\log\left(\frac{\omega_f}{\omega_i}\right) \text{.}
\end{split}
\end{equation}
The ionization fraction for Bremsstrahlung is plotted in Figure \ref{fig:brem_io}, where we have used a single type of species which the electrons interact with, namely the neutral hydrogen atoms. The dependence on $G\mu$ in the low and high $G\mu$ limits, respectively, comes from the dependence of the charged particle density on $G\mu$ which was discussed at the end of the previous section. In the high $G\mu$ limit the induced ionization fraction scales as $(G\mu)^{-3/2}$, in the small $G\mu$ limit the scaling is as $(G\mu)^{1/6}$. In the intermediate regime, the result has to be determined by numerically solving for $R_c(t)$.  The main message to draw from this calculation is the Bremsstrahlung contribution to re-ionization is negligible.

%%%%%%%%%%%%%%%%%%%%%%%%%%%%%
%%  Sychrotron Ionisation  %%
%%%%%%%%%%%%%%%%%%%%%%%%%%%%%

\subsection{Sychrotron Ionization}

Charged particles emitted from a string can be accelerated radially in the presence of primordial magnetic fields.  The resulting acceleration of the charged particles creates synchrotron radiation which can partially ionize the surrounding medium.  The spectrum of photons emitted from cosmic strings via synchrotron radiation is computed as 
\begin{equation}\
\frac{d^2n_{\gamma}(t)}{d\omega dt} \, = \, \frac{2D_e(t)e^3}{m_e}B(t)^{5/4}\left(\frac{3e}{2 m_e}\right)^{1/4}a(3/2)E(t)^{-5/4} \, \text{.}
\label{eq:spec}
\end{equation}
Here, $e$ is the electron charge, $m_e$ is the mass of the electron, $B(t)$ is the magnetic field acting on the cusp at time $t$, $a(3/2)$ is a dimensionless constant of order $10^{-1}$.  The function $D_e(t)$ comes from the flux of charged particles and is defined by
\begin{equation}
\frac{dn_e(t)}{d\Gamma} \, \equiv \,  \frac{D_e(t)}{4\pi}\Gamma^{-3/2} \ ,
\end{equation}
where $\Gamma = E(t)/m_e$.  

We can find $D_e(t)$ using equation \ref{eq:spec_g} and obtain
\begin{equation}
D_e(t) = \frac{5}{4}\pi m_e^{-1/2}\nu \mu^{1/2}t_{eq}^{1/2}t^{-7/3}\int_{t_{rec}}^{t}dt't'^{1/3}R_c(t')^{-2} \, \text{.}
\end{equation}
We then get the following expression for the ionization fraction $x_s(z)$ of synchrotron radiation:
\begin{equation}
\begin{split}
x_{s}(z) = \frac{1}{\alpha_rn_H(z)^2}\frac{8D_e(t(z))e^3}{m_e}B(t(z))^{5/4}\\
\times \left(\frac{3e}{2 m_e}\right)^{1/4}a(3/2)\left(\omega_i^{-1/4}-\omega_f^{-1/4}\right) .
\end{split}
\end{equation}

\begin{figure*}[tb]
	\centering
	\includegraphics[width = 18.0cm]{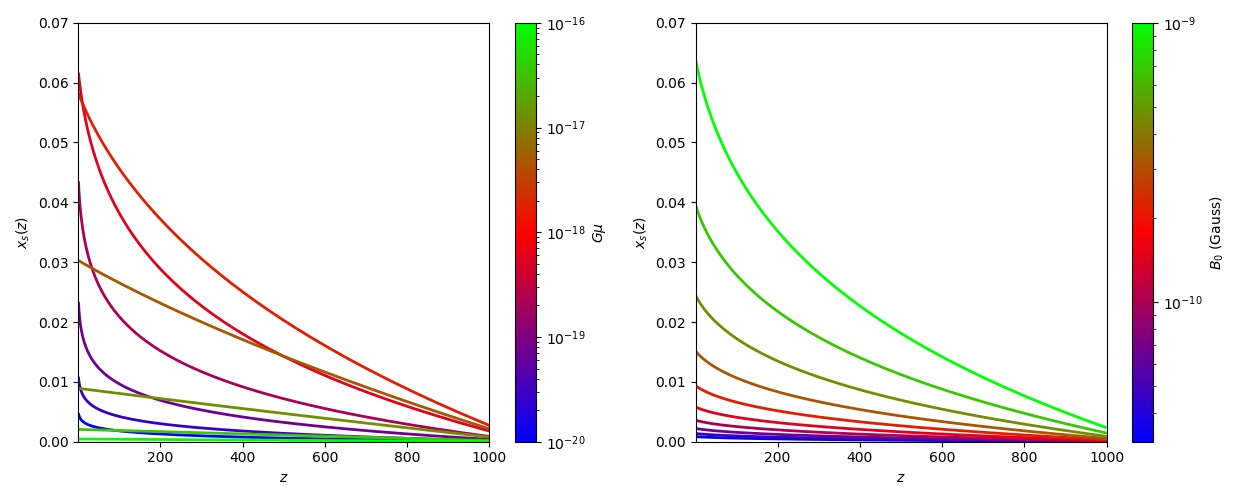}
	\caption{Ionization fraction from synchrotron radiation at different redshifts.  On the left, the value of $B_0$ is fixed at 1 nG while the values of $G\mu$ run from $10^{-24}$ to $10^{-18}$.  On the right, the value of $G\mu$ is fixed at $10^{-18}$ while the values of $B_0$ run from $10^{-12}$ Gauss to $10^{-9}$ Gauss.}
	\label{fig:sych_io}
\end{figure*}

The ionization fraction for synchrotron radiation is plotted in Figure \ref{fig:sych_io} as a function
of redshift assuming the following scaling of the primordial magnetic field
\be
B(t(z)) \, = \, B_0 (1 + z)^2 \, .
\ee
In each subplot of Figure \ref{fig:sych_io}, we explore the parameter space of the ionization fraction by fixing $G\mu$ and $B_0$ one at a time. We see that, for a large area of parameter space, the ionization fraction due to synchrotron radiation is much larger than that due to Bremsstrahlung.  As we will see in the next section, it can also dominate over the ionization fraction in the standard $\Lambda$CDM model at redshifts between recombination and reionization.

%%%%%%%%%%%%%%%%%%%%%%%%%%%%%%%%%%%%%%
%%  Effect on Reionization History  %%
%%%%%%%%%%%%%%%%%%%%%%%%%%%%%%%%%%%%%%

\section{Effect on Reionization History}

If the ionization produced after the time of recombination is important compared to what is produced in the standard $\Lambda$CDM model of cosmology, measurable effects on cosmological observables are possible. Since early ionization affects the spectrum of CMB anisotropies, constraints on new physics producing early ionization are possible. Early ionization can also affect the physics of the reionization period. We leave the study of these effects to future work, and only make a few comments.

\begin{figure*}[tb]
	\centering
	\includegraphics[width = 18.0cm]{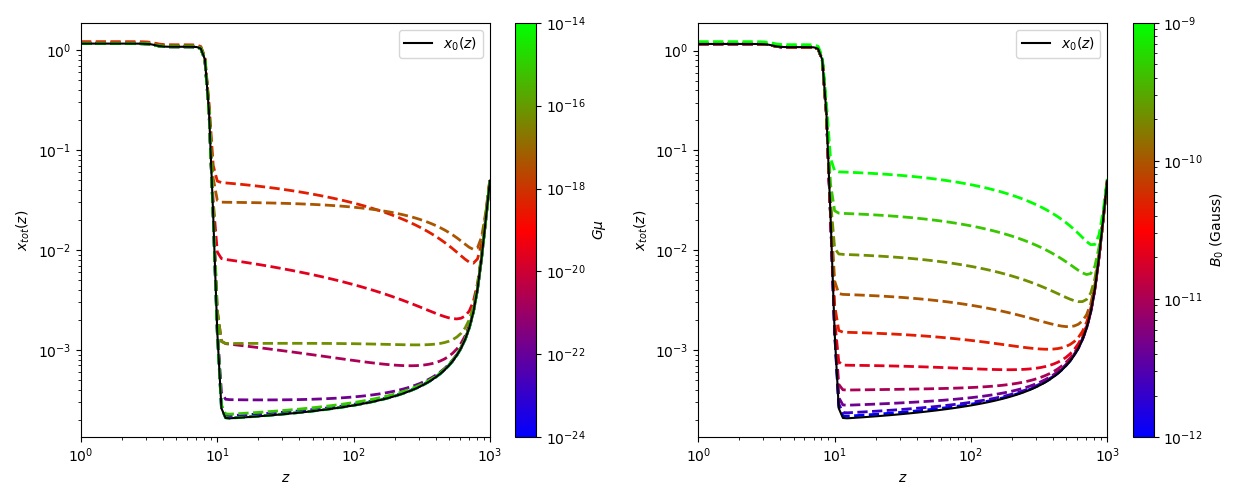}
	\caption{Ionization histories which take into account the ionization fraction from strings (dashed lines) are compared to a standard $\Lambda$CDM ionization history (black curve). On the left, the value of $B_0$ is fixed at 1 nG while the values of $G\mu$ run from $10^{-24}$ to $10^{-18}$.  On the right, the value of $G\mu$ is fixed at $10^{-18}$ while the values of $B_0$ run from $10^{-12}$ Gauss to $10^{-9}$ Gauss.}
	\label{fig:total_iofrac}
\end{figure*}

Using the CAMB code \cite{Lewis:1999bs,Lewis:2002ah} and the best fit cosmological parameters of the $\Lambda$CDM model mentioned in the introduction, we can compute the $\Lambda$CDM total background ionization $x_0(z)$ from recombination to our current time.  Given that cosmic strings emit charged particles which contribute to reionization, they contribute to the total ionization by an amount $x_{s}(z)$.  Here, we neglect Bremsstrahlung radiation since $x_{g}(z) \ll x_{s}(z)$ for the region of parameter space which interest us. The total ionization fraction
\be
x_{tot}(z) \, = \,  x_0(z) + x_{s}(z) 
\ee
takes the values described in Figure \ref{fig:total_iofrac}.

\begin{figure*}[tb]
	\centering
	\includegraphics[width = 10.0cm]{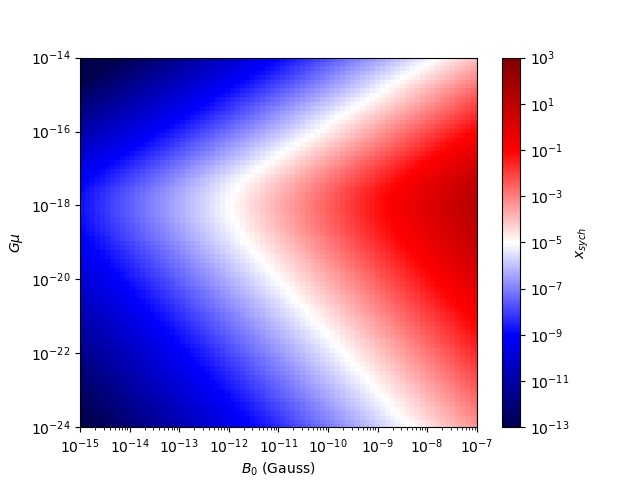}
	\caption{Color map of the ionization fraction induced by cosmic strings at $z = 12$ for different values of $G\mu$ and $B_0$.  The red region shows parameters for which synchrotron radiation from string loops dominates over the remnant $\Lambda$CDM ionization fraction.}
	\label{fig:color_map}
\end{figure*}

In Figure \ref{fig:color_map} we plot the region in the $G\mu - B_0$ parameter space for which synchrotron radiation from string loops dominates over the remnant $\Lambda$CDM ionization fraction. In this parameter space region, cosmic string loops will have a potentially measurable effect.

%%%%%%%%%%%%%%%%%%
%%  Conclusion  %%
%%%%%%%%%%%%%%%%%%

\section{Conclusion}

We have computed the contributions of Bremsstrahlung and synchrotron radiation from cosmic string cusp annihilation to the total ionization fraction of the universe in the dark ages. Whereas the contribution of Bremsstrahlung is negligible, that of synchrotron radiation can be important, depending on the values of the string tension $G\mu$ and of the primordial magnetic field $B_0$. We have identified the range of values in the $G\mu$ vs. $B_0$ parameter space where the cosmic string contribution to the ionization fraction is larger than what is predicted in the standard $\Lambda$CDM model at some point between recombination and reionization (which we have chosen to be at redshift $z = 12$ in Figure 4). In this work, we have considered non-superconducting strings. Superconducting strings lead to a larger effect, as studied in \cite{Tashiro:2012nv}.

 In work in progress, we are calculating the effects of the string-induced extra ionization of the spectrum of microwave anisotropies, with the goal of determining an exclusion region in the $G\mu$ vs. $B_0$ parameter space.

%%%%%%%%%%%%%%%%%%%%%
%% Acknowledgement %%
%%%%%%%%%%%%%%%%%%%%%

\section{Acknowledgements}
	
We would like to thank Bryce Cyr, Jonathan Sievers and Disrael da Cunha for their valuable input to the project.
RB thanks the Institute for Theoretical Physics of the ETH Zuerich for hospitality.  SL
wishes to thank NSERC for a graduate scholarship. The research at McGill is supported in part by funds from NSERC and from the Canada Research Chair program.  

%%%%%%%%%%%%%%%%%%
%% Bibliography %%
%%%%%%%%%%%%%%%%%%

\end{document}